# ESTIMATING SHEAR WAVE VELOCITY USING ARTIFICIAL NEURAL NETWORKS: A CASE STUDY OF THE TANO NORTH FIELD

*Adjei Franklin Koomson [1*], Afari Priscilla Sitsofe [1,] Adjei Osae Teddy [2]*
*Azubi Africa [1*]*
franklin.adjei@azubiafrica.org [1*], priscilla_afari@yahoo.com [1], taosae17@gmail.com [2]

## ABSTRACT

Shear wave velocity is an important parameter for determining lithology, porosity and the dynamic properties in geo-mechanical studies. However, due to time and cost limitations, shear wave velocity is not available at all intervals and in all wells. In this paper, well logs with strong correlation to shear wave velocity were determined and used to predict the shear wave velocity for the Tano North Field. Four different methods were used to estimate the shear wave velocity under three different conditions. Then, based on obtained coefficient of determination and average absolute percent relative error between real and predicted values of shear wave velocity, the final results were compared. The results of this work demonstrated that the neural network based on multiple variables can estimate the shear wave velocity better than other methods us

## INTRODUCTION

### 1.1 GENERAL INTRODUCTION

The complex nature of oil and gas reservoirs is a challenge in the petroleum industry. The unavailability of reliable information makes prediction of reservoir parameters very difficult. Knowledge of dynamic properties of reservoir rocks such as the Young's modulus, Poisson ratio and Shear modulus is indispensable to geo-mechanical studies of reservoirs, as they help in analysis of well characteristics such as well bore stability, fracturing design, geo-mechanical modelling and wellbore trajectory design. Shear wave velocity is used to calculate the dynamic modulus of the formation. It could also be used in further studies of the petro-physical properties of the field. Shear wave velocity information, however, is not readily available in wells due to high acquisition costs and other limitations. (Zaboli et al. 2016).

In years past, the use of classical data processing tools was enough to solve met geological problems, (Eskandari H. et al, 2004). However, more complex reservoirs are being discovered, and these tools have proved insufficient in solving these problems (Akhundi et al 2014). Shear waves are elastic body waves that are propagated perpendicular to the movement of particles. This mode of propagation makes them move slower than compressional waves and hence a wider scope of measurement. In recent years, the Dipole Shear Sonic Imager (DSI), a relatively novel addition to the petroleum industry, has been used to measure shear wave velocity directly. One severe limitation of the use of this tool, however, is its cost. Also, because of the novelty of the tool, oldest wells do not have any recorded shear wave data. Due to this, numerous methods have been presented to estimate shear wave velocity from other well log data that are recorded in most wells.

In petro-physics, the methods used to study mechanical rock properties can be put into three broad categories; statistical and empirical methods, laboratory tests and theoretical studies (Eskandari H et al, 2004). Laboratory tests (direct methods) give exact values of rock properties, but these tests take a lot of time to produce results, and they are very costly. Conventional well logging methods are also used to ascertain certain petro-physical properties. Compressional wave velocity data is much more readily accessible in traditional well logs, as compared to shear wave data (F. Hadi et al, 2018). Also, there is very little accessible data on the measurement of shear wave velocity in the laboratory. This is mainly because shear wave measurements cannot be taken at very low pressures. Since shear wave velocity depends on properties such as lithology and loading conditions, laboratory measurements often come up inaccurate, because reservoir conditions cannot be properly simulated in the laboratory. Statistical predictions are highly dependent upon the amount of data collected. Such predictions may also be used for well planning. However, most previous relationships have been developed from limited core measurements and very few attempt to predict the Vs of a field case.(F. Hadi et al, 2018)

Artificial Neural Networks (ANNs) have been, and are being used to model complex reservoirs due to their ability to relate unknown parameters. The fundamental basis of ANNs is their ability to learn and generalize the behavior of a system using sets of connection weights. Conventional well logs will be used to estimate shear wave velocity in wells, with the help of artificial neural networks.

## 1.2 PROBLEM STATEMENT

Various well logs bear correlation with shear wave velocity, and can be used in its estimation. However, the methods available for shear wave velocity determination are time consuming, expensive, require technical know-how and are not even reliable due to limited data from cores. (Hadi F.A. et al, 2018). For this reason, we will focus on using Back Propagation Neural networks (with the MATLAB as an interface) in the estimation of shear wave velocity from well logs for a reservoir in the North Tano field (Expanded Shallow Water Tano Block, Tano - Cape three-point basin) in Ghana.

### 1.3 OBJECTIVES

### 1.3.1. Main Objectives

The main objective is to find the correlation between various conventional well longs and shear wave velocity using the Back Propagation Neural Networks, and then use these logs to estimate shear wave velocity.

### 1.3.2. Specific Objectives

1. To ascertain the relationship between various well logs and shear wave.
2. To determine which well logs are appropriate for use in the estimation of shear wave velocity.
3. To show how BPNN can be used to estimate shear wave velocity from appropriate well logs using MATLAB as the programming language.
4. To evaluate and compare the result from this method of estimating shear wave velocity.
5. To ascertain the validity or otherwise of the us of ANNs in shear wave prediction.

### 1.4 JUSTIFICATION

The methods used to estimate shear wave velocity can be put into three broad categories; statistical and empirical methods, laboratory tests and theoretical studies. ("F.A. Hadi" 2018). This makes it very relevant for the industry to estimate it from well logs since a correlation exists between the well logs and shear wave velocity. The remaining methods do not give more concise estimation of shear wave as compared as estimating it from well logs. BPNN helps to give required output from the output neurons depending on the type of data fed to the input neurons. Learning, understanding and writing of MATLAB codes is very easy. Also, MATLAB simplifies Deep Learning matrices and makes them easier to understand.(Kim, 2017.).

### 1.5 LOCATION AND GEOLOGY OF THE STUDY AREA

The region of the Tano basin involved in this study is the northern part called North Tano field. It is one of the three discoveries of the 'Expanded Shallow Water Tano Block'. The combined surface area of the block is 1508 $km^2$, and the operator company is Erin Energy Ghana Limited (60% interest). Other contracting parties includes GNPC EXPLORCO (10% interest), Base Energy Limited (20% interest) and GNPC (10% interest). North Tano field is located in the south-western part of Gulf of Guinea. This portion of the basin was discovered by Philips Petroleum in 1980 and promises a high yield of hydrocarbons. It is located about 15km offshore Ghana. The depth of the water is about 55 m. The Tano forms the eastern extension of the Deep Ivorian basin where most of the hydrocarbons are produced. The other two fields in the block are South and West Tano discoveries.

The Tano North field is located in the Tano basin. The deposition of rock in the Tano basin began during the Aptian age of the early cretaceous era. Simultaneously, during this period, there was a continental rift between the South America plate and the African plate, which resulted in the formation of the Tano basin. Also, during this same time, the Atlantic Ocean in the Albian opened, resulting in the deposition of rich organic matter in the Turonian and the Cenomaian. The type of organic matter deposited in the Tano basin is mainly Cretaceous, with the source rocks being Turonian-Cenomanian shales, and the reservoir rocks being Albian sandstone. The traps are both structural and stratigraphic. The Tano basin forms an indelible part of the West African transform margin, and is found between the Romanche and St. Paul
faults (Baik et al, 2000).

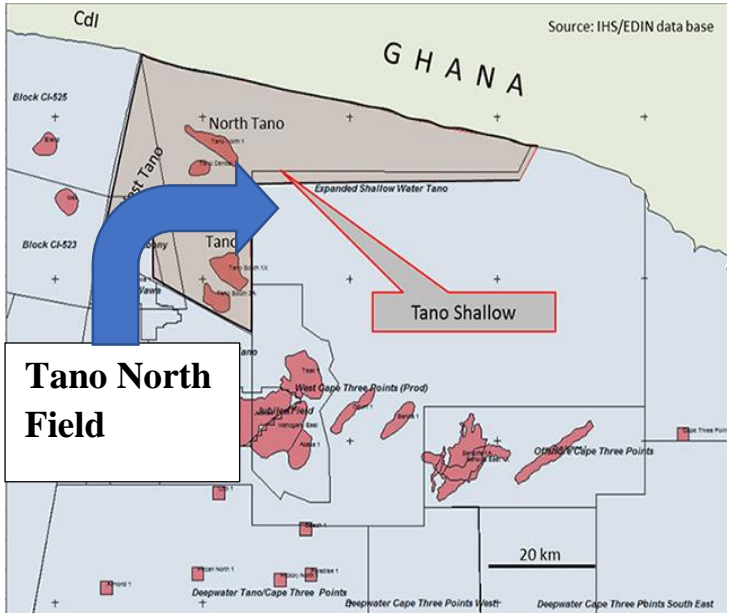

Figure 1.1: Location of North Tano field

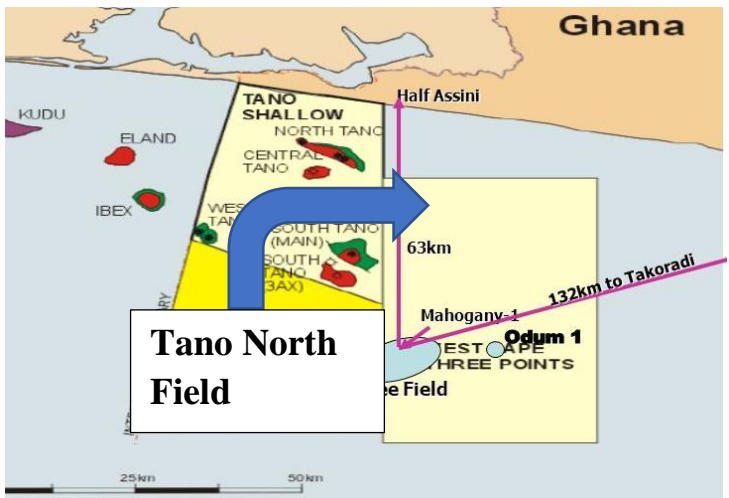

Figure 1.2: Distance of North Tano Field from onshore

## LITERATURE REVIEW

Seismic exploration is the most important part of oil and gas exploration. Seismic exploration is basically the study of the earth and its shape, the speed at which waves travel through it and other physical properties. The speed of waves through a rock depends on certain characteristics of the rock such as the tensile and compressive strength. This implies that the properties that affect wave velocity in rocks are mostly intrinsic. Examples of these properties include porosity, hardness and rock type (Zaboli *et al.* 2016).

## 2.1 SHEAR WAVE VELOCITY

Seismic waves are waves generated either naturally by earthquakes or other means such as explosions (Shearer, 1999). The petroleum industry has adopted the use of seismic waves as a way of exploration for oil and gas. Explosions are generated by explosive devices such as dynamite, and these explosions send seismic waves through the earth. Sensors are placed on the surface in trucks to record the responses generated by the explosions, and these responses are analyzed for signs of hydrocarbon existence. Another way of generating seismic waves is by using the vibrator or air gun. The vibrator is used onshore whereas the air gun is used on the sea. The vibrator generates seismic waves by having direct impact on the ground. The air gun sends sound waves into the water (Tixier, Algier, & Doh, 1999).

Four different types of seismic waves are generated through the earth at any point; shear waves (S-wave), compressional waves (P-waves), Rayleigh waves and Love waves (Moon, Ng, and Ku 2017). These waves can be placed in two broad categories, that is, surface and body waves. Rayleigh and love waves fall under the category of surface waves. They are, however, not significant to the petroleum industry (Moon, Ng, and Ku 2017). P-waves and S-waves, are, however, of significance in the industry because they travel into the earth.

P-waves are longitudinal waves which are propagated in the direction in which the force is travelling. Since the earth is incompressible, the wave moves very fast through it. The S-wave, on the contrary, is propagated transversely to the direction of the force. S-waves are generally slower than P-waves, and they do not travel through fluids as the P-waves do (Petcher, Burrows, and Dixon 2014). Because they have a polarization, S-waves can be significantly affected by anisotropies of the medium (i.e. a medium whose properties differ in different orientations). If one orientation is much faster than the other ones, then it may be possible to detect this by observing different propagation velocities in the different polarization planes (Eskandari, H.; Rezaee, M.R; Mohammadnia 2004) This aspect underpins efforts to identify open fractures through shear wave variations. The direction of propagation of both P and S wave are illustrated in the figure below.

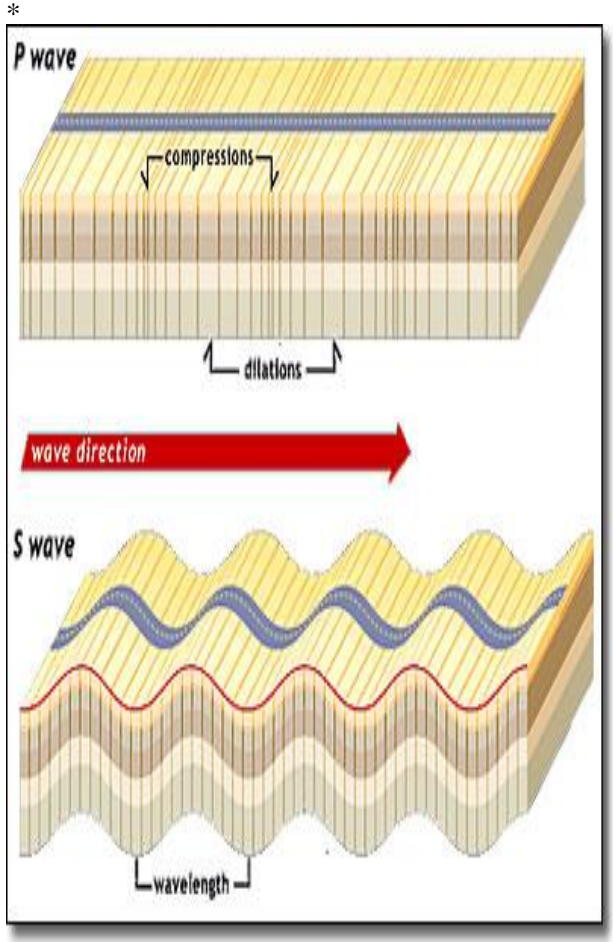

*

Figure 2.1: Movement of waves through the earth

Seismic receivers in the marine environment are called hydrophones (these detect pressure changes caused by compressional waves, converting them to electrical signals. A single streamer of 16-24 equally-spaced hydrophones (channels) is common, but nowadays, multiple parallel streamers are used, producing many 100's of channels of data. In the onshore environment, seismic receivers are called geophones. Geophones are clamped to the ground and detect motion through vibrations of a coil of wire moving through a magnetic field, producing time-varying electrical signals (similar to the operation of standard microphones, as used in telephones). Measurement of S-wave velocity on the field directly is very rare and very expensive. One such tool is the Dipole Shear Sonic Tool. The disadvantage of using such a tool is that it requires one or more boreholes to be drilled into the formation (Petcher, *et al* 2014)

Many methods have been devised to ascertain the velocity of shear waves, both in the field and in the laboratory. Examples are seismic reflection and refraction survey, downhole and cross holed methods and suspension logging. (Greenwood 2015). All these methods are invasive methods, that is, they require the use of down-hole equipment to be determined. This paper, however, explores the use of statistical methods to find shear wave velocity.

Shear waves have been correlated to many geotechnical parameters. These include peak friction angle, undrained shear strength, soil unit weight, lateral earth pressure coefficient, compressibility, porosity or void ratio, degree of consolidation, stress history and degree of saturation (Moon, Ng, and Ku 2017). (2.1) to (2.5), show the importance of wave velocity in calculating the dynamic modulus.

$$E = 1.34 * 10^{10} \rho V_s^2 \left(\frac{3V_p^2 - 4V_s^2}{V_p^2 - V_s^2}\right)$$
(2.1)

$$K = 1.34 * 10^{10} \rho \left(\frac{3V_p^2 - 4V_s^2}{3}\right)$$
(2.2)

$$G = 1.34 * 10^{10} \rho V_s^2$$
(2.3)

$$C = \frac{1}{K}$$
(2.4)

$$\vartheta = \frac{V_p^2 - 2V_s^{\mu2}}{2V_p^2 - 2V_s^2}$$

(2.5)

Where, E is Young's modulus in Psi, K is bulk modulus in Psi, G is shear modulus in Psi, C is compressibility modulus in Psi, $\vartheta$ is Poisson's ratio, $V_p$ is compressional wave velocity in ft/µs, $V_s$ is shear wave velocity in ft/µs and $\rho$ is density in g/cm is density in g/cm$^3$. (Zaboli et al. 2016)

(2.6) shows the relationship between material mass density and the constrained modulus. (2.7) shows the relationship between shear wave velocity, material mass density and shear modulus. The wave velocities are related to each other in (2.8) through the Poisson's ratio.

$$Vp = \sqrt{M/\rho}$$

(2.6)

$$Vs = \sqrt{G/\rho}$$

(2.7)

$$\frac{Vp}{Vs} = \sqrt{\frac{1-v}{0.5-v}}$$

(2.8)

(Greenwood 2015)

One main reason for determination of shear wave velocity is to learn more about the geological and geotechnical features of the reservoir. It is also used to ascertain anisotropy in rocks, determine fracture strike and in $CO_2$ injection monitoring. (Greenwood 2015). The ratio of compressional to shear wave velocities give information on lithology, porosity, and the dispersion of waves, which yields useful permeability and fracture data. (Petcher, Burrows, and Dixon 2014).They can also be used to access the amount of gas hydrates in a sedimentary sequence (Moon, Ng, and Ku 2017). This paper focuses on the correlation between shear stress and available data on geophysical properties such as porosity and density as a function of the various well logs.

## 2.2 PETROPHYSICAL PROPERTIES

One of the goals of the petroleum industry is to create a reliable model to predict shear wave velocity in wells using traditional well logs such porosity logs, grain density, cementation exponent, just to name a few (Akhundi, Ghafoori, and Lashkaripour 2014).

Compressional velocity is can be measured with the acoustic log, and the clay volume fraction can be measured with the gamma ray log (Castagna, Batzle, and Eastwood 1984). Porosity can be derived from sonic, density and neutron logs. Permeability can be determined by use of the irreducible water saturation at a constant bulk volume of water and may also be determined using NMR (Nuclear Magnetic Resonance) log (Rezaee, R.; Kadkhodaie Ilkhchi, A.; Barabadi 2007). The cementation exponent can be determined by combining log-porosity data and resistivity data using the Pickett plot (Petcher, Burrows, and Dixon 2014).

### 2.2.1. Gamma Ray Log
High energy electromagnetic rays that are emitted spontaneously from a radioactive element are referred to as gamma rays. The earth emits gamma rays, and majority of those waves are emitted by the radioactive isotopes potassium ($^{40}$K), uranium and thorium. Potassium is the most abundant, and therefore emits the most waves. These gamma rays can be measured by the Gamma Ray Logging tool, which is a passive tool. When gamma rays pass through the formation, they collide with the atoms of the particles of the formation, and are scattered. This phenomenon is known as *Compton scattering*. After the collision, the gamma rays lose energy, and are absorbed by the atoms of the formation, by *photoelectric effect*. The gamma ray logging tool measures the weight concentrations of the gamma rays after the scattering, and corrections for borehole size and other parameters are made. The measurement made by the tool corresponds the following equation:

$$GR = \frac{(\rho i V i A i)}{\rho b}$$

(2.9)

where

$\rho i$ are the densities of the radioactive minerals,

$V i$ are the bulk volume factors of the minerals

$A i$ are proportionality factors corresponding to the radioactivity of the mineral,

Most reservoir have simple lithologies consisting of layers of sandstone and shales, or shales and carbonates. The volume of shale, $V_{sh}$, of the formation can be calculated once these lithologies have been identified. The volume of shale is important for the discrimination between reservoir and non-reservoir rock. To calculate $V_{sh}$, the gamma ray index is first calculated using the relation:

$$I_{GR} = \frac{GR_{log} - GR_{min}}{GR_{max} - GR_{min}}$$

(2.10)

Where $GR_{log}$ = the gamma ray reading at the depth of interest

$GR_{min}$ = the minimum gamma ray reading. (Usually the mean minimum through a clean sandstone or carbonate formation.)

$GR_{max}$ = the maximum gamma ray reading. (Usually the mean maximum through a shale or clay formation.) (Bowen *et al* 2003)

### 2.2.2. Sonic Logs

The sonic (or acoustic) log is a kind of well logging methods which measures the properties of elastic waves as they pass through the formation. The propagation parameters of sonic wave through formation including: velocity, amplitude and frequency dispersion These propagation parameters are a reflection of the elastic mechanical properties which is related to density (or lithology, porosity, and fluid contents) of the formation

Different methods are developed:

- Sonic velocity log

- Sonic amplitude log

- Long spaced sonic log

- Array Sonic log

**Compressional wave velocity (Vp)**

$$V_p = \sqrt{\frac{E}{\rho} * \frac{(1-v)}{(1+v)(1-2v)}} = \sqrt{\frac{K + \frac{4\mu}{3}}{\rho}} = \sqrt{\frac{\lambda + 2\mu}{\rho}}$$

(2.11)

Where K is the bulk modulus (the modulus of incompressibility),

$\mu$ is the shear modulus (modulus of rigidity, also called the second Lamé parameter),

ρ is the density of the material through which the wave propagates

λ is the first Lamé parameter

ν is the poisson's ratio

Density shows the least variation out of these parameters. As a result, velocity is mostly controlled by the bulk modulus (K) and shear modulus (μ). The amount of time a wave takes to travel through the formation to a certain distance is known as the transit time. It is the reciprocal of velocity, and is measured in μs/ft. Wave velocity can also be determined from well logs using the formula $V_p = \frac{304.8}{DT}$, where DT is the sonic log. (Akhundi, Ghafoori, and Lashkaripour 2014)

**Shear wave velocity**

$$V_s = \sqrt{\frac{E}{\rho} * \frac{1}{2(1+v)}} = \sqrt{\frac{\mu}{\rho}}$$
(2.12)

μ = 0 for fluids, so shear wave cannot propagate in liquids.

**Poisons ratio**

$$\frac{V_p}{V_s} = \sqrt{\frac{2(1-v)}{1-2v}} = \frac{K}{\mu} + \frac{4}{3}$$
(2.13)

This ratio always >1 since K and μ are positive, and that means $V_p$ is always larger then $V_s$.

**Porosity Determination (The Wyllie Time Average Equation)**

The velocity of elastic waves through a given lithology depends on the porosity of the formation. Wyllie proposed a simple equation to describe this relationship and it was named the time average equation.

$$\frac{1}{V} = \frac{\emptyset}{V_p} + \frac{1-\emptyset}{V_{ma}}$$
(2.14)

$$\Delta t = \emptyset \Delta t_p + (1 - \emptyset)\Delta t_{ma}$$
(2.15)

Hence, $\emptyset_s = \frac{\Delta t - \Delta t_{ma}}{\Delta t_p - \Delta t_{ma}}$
(2.16)

(Bowen *et al* 2003); (O.Serra, 1984); (Bateman, 1985)

### 2.2.3. Neutron Logs

When neutrons move through the formation, they collide with the particles of the formation, and energy is lost with each collision. Eventually, the neutrons are slowed down and captured by the nuclei of the atoms. Neutrons lose the most energy when they collide with hydrogen atoms. The captured nuclei then get excited and emit gamma rays. The neutron logging tool is an active tool which bombards the formation with high energy neutrons. The number of neutrons that reach the receiver of the tool depends on the hydrogen index (amount of hydrogen) in the formation. A high hydrogen index indicates the presence of fluids (oil or water) in the formation. Limestone porosity is the calibration factor for hydrogen index. Either captured or non-captured neutrons are recorded, depending on the type of logging tool used. For this log, the porosity of the formation is determined based on the neutrons captured by the formation.

**Theoretical equation**

$$\emptyset_N = \emptyset S_{XO}\emptyset_{Nmf} + \emptyset(1 - S_{XO})\emptyset_{Nhc} + V_{sh}\emptyset_{sh} + (1 - \emptyset - V_{sh})\emptyset_{Nma}$$
(2.17)

Where,

$\phi_N$  =  Recorded parameter

$\phi\, S_{xo}\, \phi_{Nmf}$  =  Mud filtrate portion

$\phi\, (1 - S_{xo})\, \phi_{Nhc}$ =  Hydrocarbon portion

$V_{sh}\, \phi_{Nsh}$  =  Shale portion

$(1 - \phi - V_{sh})\, \phi_{Nhc}$  =  Matrix portion

$\phi$  =  True porosity of rock

$\phi_N$  =  Porosity from neutron log measurement, fraction

$\phi_{Nma}$  =  Porosity of matrix fraction

$\phi_{Nhc}$  =  Porosity of formation saturated with hydrocarbon fluid, fraction

$\phi_{Nmf}$  =  Porosity saturated with mud filtrate, fraction

$V_{sh}$  =  Volume of shale, fraction

$S_{xo}$  =  Mud filtrate saturation in zone invaded by mud filtrate, fraction

(Bowen *et al* 2003)

## 2.2.4. Formation Density Logs

The formation density log measures the overall porosity of the formation. It is also used in the detection of gas bearing formations and evaporates. Formation density tools are active tools which bombard the formation with radiation, and measure the amount of radiation that comes back to the sensor. A new form of formation density log, called the litho-density log, has added features that make it more effective. It has short spaced and long spaced detectors as the FDC tool, and a caesium-137 source emitting gamma rays at 0.662 MeV. These detectors are more efficient, as they can detect rays with as high energy as 0.25 to 0.662 MeV, and as low energies as 0.04 to 0.0 MeV. For a molecule made up of several atoms, a photoelectric absorption cross section index, Pe, may be determined based upon atomic fractions. Thus Pe= ($AiZiPi)/($AiPi).

### 2.2.4.1 Determination of Porosity

The porosity f of a formation can be determined from the bulk density if the mean density of the rock matrix and that of the fluids it contains are known. The bulk density $\rho_b$ of a formation can be expressed as a linear contribution of the density of the rock matrix $\rho_{ma}$ and the fluid density $\rho_f$, with each present is proportions $(1- \varphi)$ and $\varphi$ , respectively :

$$\rho b = (1 - \varphi)\,\rho ma + \varphi\,\rho f \quad (2.18)$$

When solved for porosity, we get

$$\emptyset = \frac{\rho_{ma} - \rho_b}{\rho_{ma} - \rho_f} \quad (2.19)$$

$\rho_b$ = the bulk density of the formation

$\rho_{ma}$ = the density of the rock matrix

$\rho_f$ = the density of the fluids occupying the porosity

$\varphi$ = the porosity of the rock.  (Bowen *et al* 2003)

Petro-physical properties affect shear wave velocity. Since these properties are obtained from conventional well logs, there exists a relationship between those conventional well logs and shear wave velocity. Shear wave velocity helps to develop a basis for predicting petro-physical information from seismic data. It also makes common hindrances in Petroleum Engineering such as subsidence become easier to analyze and overcome. (Castagna et al, 1997) *.

## 2.3 ARTIFICIAL NEURAL NETWORK

An artificial neural network is a computer model that mimics simple biological learning processes, in that it simulates specific functions of natural neurons in human nervous system. ANNs have been used in the field for automating tasks in seismic processing. Such tasks include trace editing, which requires a lot of manual interpretation.(Ali, 1994). The network develops a relationship between the input it is fed, and their corresponding outputs, by assigning weights to the inputs according to their correlation to the output parameters (Lim, 2005). The ANN, unlike more conventional methods which have fixed algorithms for problem solving, uses a form of learning in which it performs a non-linear mapping between the input and the output data. This helps the network gather more information about the problem at hand (Caldero et al, 2000).

The network performs the mapping based on the inputs and outputs fed into it by the interpreter. As the network learns and is able to successfully reproduce the recognized pattern, it can be used in the prediction of new data. It can also be used in identification of patterns and mapping problems. (Sahin, Guner, and Ozturk 2016)

A neural network can familiarize itself with complex non-linear relationships, even when the input information given is less precise and noisy. Also, no prior knowledge of input parameters is needed by the network as to the nature of the correlation between the input and output parameters, as is in the case of other statistical and empirical methods. This gives the neural network a marked advantage over the other empirical and statistical methods. Neural networks are well suited for problem solving in the industry because they are able to recognize patterns in given data even with noise, and they have other abilities like market forecasting and process modelling. They can also determine underlying relationships between input and outputs (Akhundi, Ghafoori, and Lashkaripour 2014)

### 2.3.1. STRUCTURE OF ANNs

ANNs consist of three layers of neurons. These neurons are organized in layers, with each layer performing a specific task. The input layer receives the input data or information and sends it over to the middle layer. The middle layer is responsible for the analysis of the entered data. The output layer translates the processed data into a more understandable form and displays it to the user. The network utilizes a nonlinear tangent sigmoid function (tansig) for the transfer of data from the input to the middle layer, and a purlin function for the transfer of data from the middle to the output layer (Akhundi, Ghafoori, and Lashkaripour 2014). ANNs need to be trained in order to be able to predict data sets. This training is necessary for the network to be able to properly perform its task properly. For this stage, the network is provided with inputs and their corresponding outputs. This process is also termed as learning (RBC Gharbi et al, 2005)The network is repeatedly exposed to these data sets, allowing it to be able to study the correlations and assign the various weights to each of the input parameters provided. An input parameter can be made equal to one and called a bias. This would make up for the effects that are not accounted for by the input parameters. Also, there is a bias neuron in the hidden layer. This neuron is not connected to any neurons in the previous layer, but connected to all neurons in the next layer (Asadi 2017).

Feed-forward networks generally consist of one or more hidden layers, followed by one output layer of neurons. (Asadi 2017). The network uses a Levenberg-Marquardt learning rule to modify the weights assigned to the neurons by repeated iterations of the function to obtain the best relationship between the input and output parameters (Rezaee et al, 2007).

### 2.3.2. ERROR MINIMIZATION THROUGH LEARNING

The back propagation method is the most widely used training method for ANNs. It produces a result that has a least square fit relationship with the desired output. It does this by finding a gradient in terms of network weights. BPNNs consist of two phases; the forward and backward phases. For the forward phase, the network receives the input and is fed forward until a prediction is generated. (Asadi 2017). For the backward phase, the error signal is back propagated into the network through the output layer and the weight adjustments are calculated using a mathematical correlation that minimizes the sum of squared errors (Rezaee, Ilkhchi, and Barabadi 2007). Weights are calculated by reducing the difference between the network outputs, after one set of inputs have been propagated through the network.

### 2.3.3 GENERALIZED DELTA RULE

The delta rule adjusts the weight as follows:

Assuming that the input node contributes to the error of the output node, the weight between the two nodes is adjusted proportionate to the input value, $x_j$ and the output error, $e_i$.

$W_{ij} = w_{ij} + \alpha * \phi_i * x_j$

But, $\phi = \varphi' * v_i * e_i$

Where

$\alpha$ = learning rate

$W_{ij}$ = new weight

$w_{ij}$ = old weight

ei = The error of the output node i

vi = The weighted sum of the output node i

$\varphi'$ = the derivative of the activation function $\varphi$ of the output node i.

The network can subjected to additional training if the input functions are adjusted in a new situation or circumstance (RBC Gharbi et al, 2005).

BPNNs can be used as an alternative tool for modelling processes which rely on significant well logging data for identification of relations. BPNNs provide the flexibility and adaptability needed of a model that can estimate shear wave velocity, which is very much affected by many random variables. The network can make up for the lack of linearity and the variable behavior of formations, even when these variables are often unknown. (Sahin et al, 2016).

# METHODOLOGY

## 3.1. DATA COLLECTION AND CONDITIONING

- Ms. Excel was used to display data and interpret by generating graphs of depth against each well log.
- Null and/or inconsistent values were deleted.
- SPSS was used to refill the gaps by interpolation

## 3.2. GENERATION OF SHEAR WAVE VELOCITY DATA

- Using the Castagna Equation, Shear wave velocity data were generated from the sonic log.

$$Vs = 0.80416 * Vp - 0.85588$$

Vp is the inverse of the sonic log reading

### 3.3. NEURAL NETWORK DESIGN

Neural networks are generally designed in six steps as shown in the diagram below:

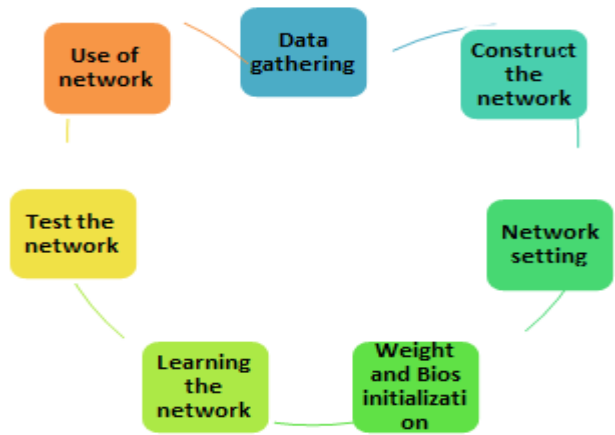

**Figure 3.1: Illustration of the mode of operation of Artificial Neural Networks**

For this method, the ANN tool available in the MATLAB software was used. The network is going to be constructed first with dummy data, in order to test the network and initiate the learning network. After he learning is done, the network will be tested with actual data. In practice, more than 70% of the data available is used in the learning process, and only 15% is used in the actual testing. (Zaboli et al. 2016)

For this research work, well log data will be used to estimate shear wave velocity. Several methods would be used, and these methods would be compared to ascertain the best method. For each method, after generating the shear wave velocity values, and predicting the values through the various methods, the measure of the closeness of the predicted and generated values, that is the Coefficient of determination ($R^2$) and average absolute percent relative error (AAPRE) between real and predicted values of shear wave velocity are calculated. The best method would be selected and recommended after all these tests are done. The methods to be evaluated includes; Linear Regression, Multiple Linear Regression and Artificial Neural Networks (single and multiple variable).

$R^2$ is an indication of the match of the generated result to the equation, and shows the validity of the model or equation. If $R^2$ is closer to 1, it means the real and predicted values are fitted. The AAPRE is a measure of the relative absolute deviation from the real values. These are determined as follows:

$$R^2 = 1 - \frac{\sum_{i=1}^{n}(x_i - \hat{x}_i)^2}{\sum_{i=1}^{n}(x_i - m_x)^2}$$

(3.1)

$$AAPRE = \frac{\sum_{i=1}^{n}\frac{|x_i - \hat{x}_i|}{x_i} * 100}{n}$$

(3.2)

Where $x_i$ is real value, $\hat{x}_i$ is predicted value, $m_x$ is average of real values and n is number of data. (Akhundi, Ghafoori, and Lashkaripour 2014; Zaboli et al. 2016)

The network was used to determine the $R^2$ and AAPRE:

- For Resistivity Data against shear wave velocity data.
- For Depth, GR, NHPI and RHOB against shear wave velocity data.
- For an interval of the same well whose shear wave velocity data was assumed to be unknown.
- For an interval of a different well in same field whose shear wave velocity data is assumed to be unknown.

### 3.4. LINEAR REGRESSION

Regression is a statistical method used to estimate a mathematical correlation to determine an unknown value using known values.(Complete Dissertations,2013) A linear equation for the correlation of NPHI or RHOB or GR or Depth and shear wave velocity was obtained. This linear equation is as follows;

$$V_s = A_0 + A_1(NPHI \text{ or } RHOB \text{ or } GR \text{ or } Depth)$$
(3.3)

- The linear regression was performed using SPSS (IBM statistics 25) software.
- The $R^2$ and AAPRE were generated from MATLAB and Ms. Excel respectively and recorded

### .3.5 MULTIPLE LINEAR REGRESSION

For this case study, shear wave velocity was predicted using conventional well logs such as depth, the neutron porosity logs (NPHI), bulk density logs (RHOB) and Gamma Ray (GR) logs. This was also done using the SPSS software. This established a relationship between the input parameters (NPHI, RHOB, GR and Depth) and Shear Wave velocity. From this relationship, the coefficients- $a, b, c$ and $d$ - were determined and used in the equation

$$V_s = a + bDepth + cNPHI + dRHOB + eGR$$
(3.4)

- The multiple linear regression was performed using SPSS (IBM statistics 25) software.
- The $R^2$ and AAPRE were generated from MATLAB and Ms. Excel respectively and recorded

# RESULTS AND DISCUSSION

## 4.1. SELECTION OF APPROPRIATE DATA

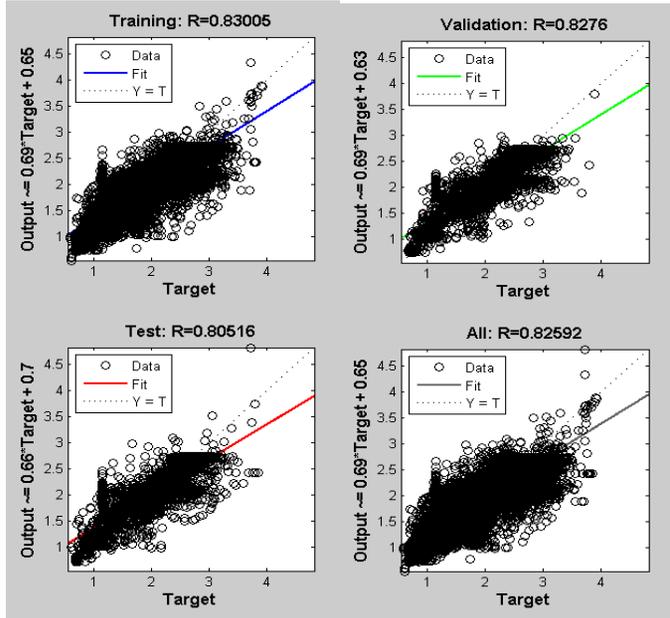

Figure 4.1: Regression plots for neutron and density logs as inputs

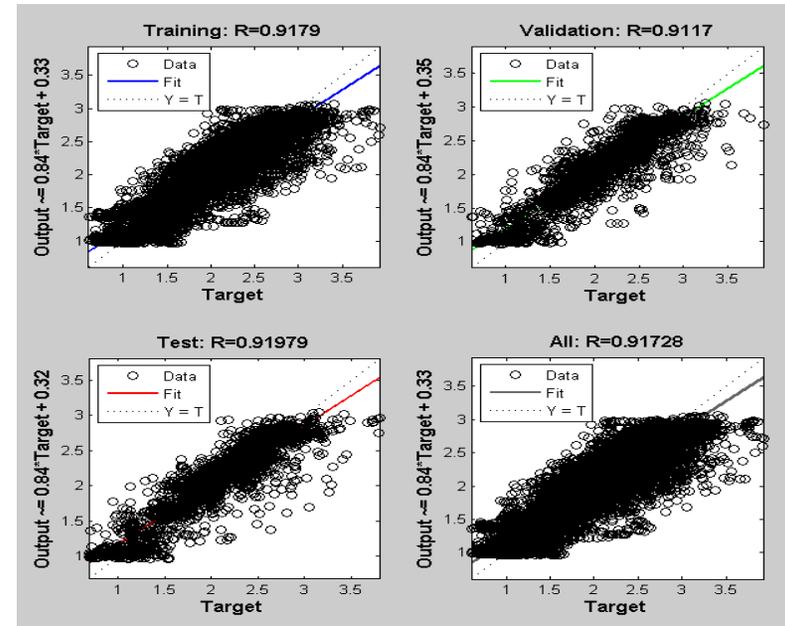

Figure 4.2: Regression plots for depth and gamma ray logs

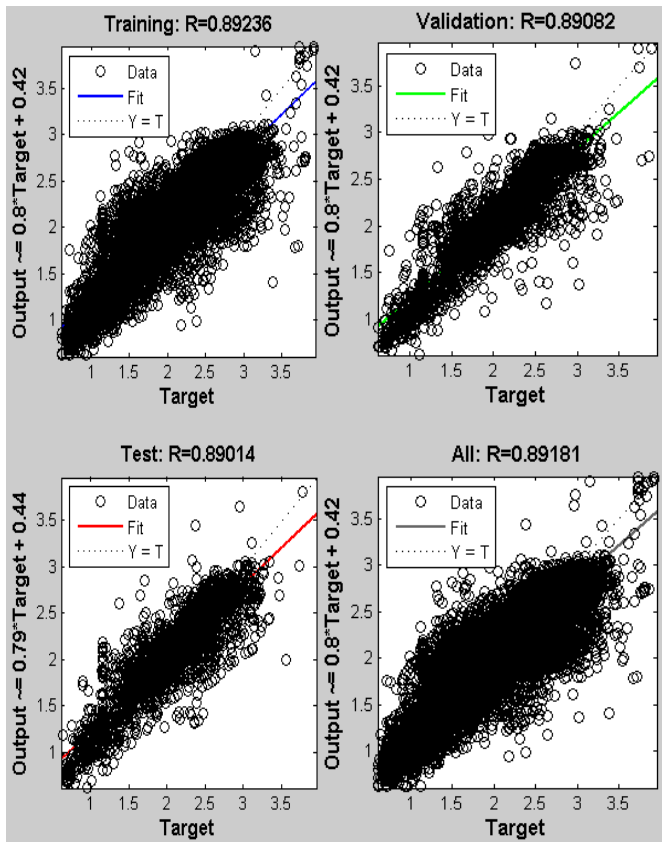

Figure 4.3: Regression plot for neutron, density and gamma ray logs

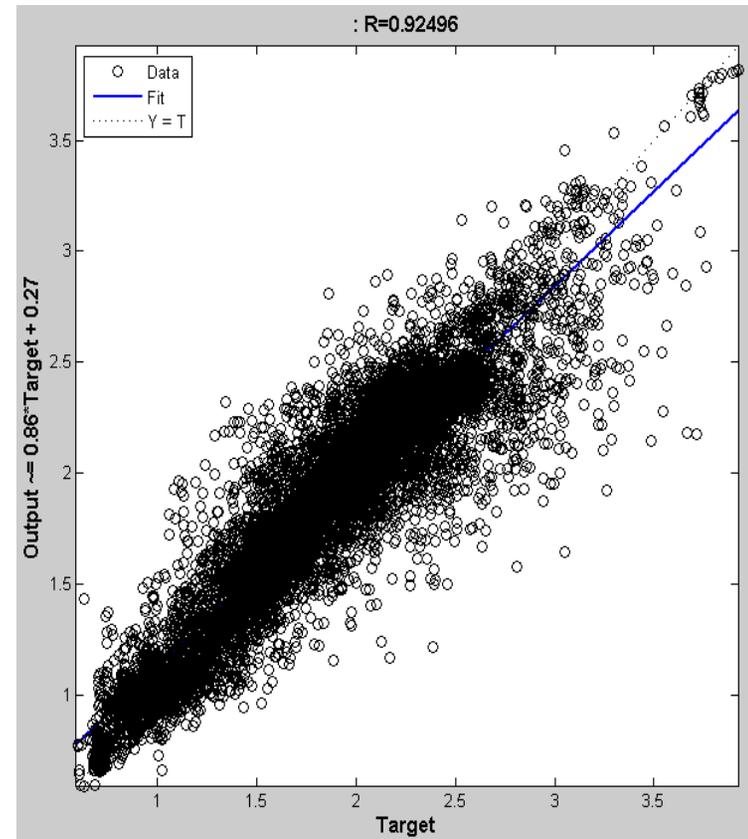

**Figure 4.4: Regression plot for porosity and GR logs and Depth**

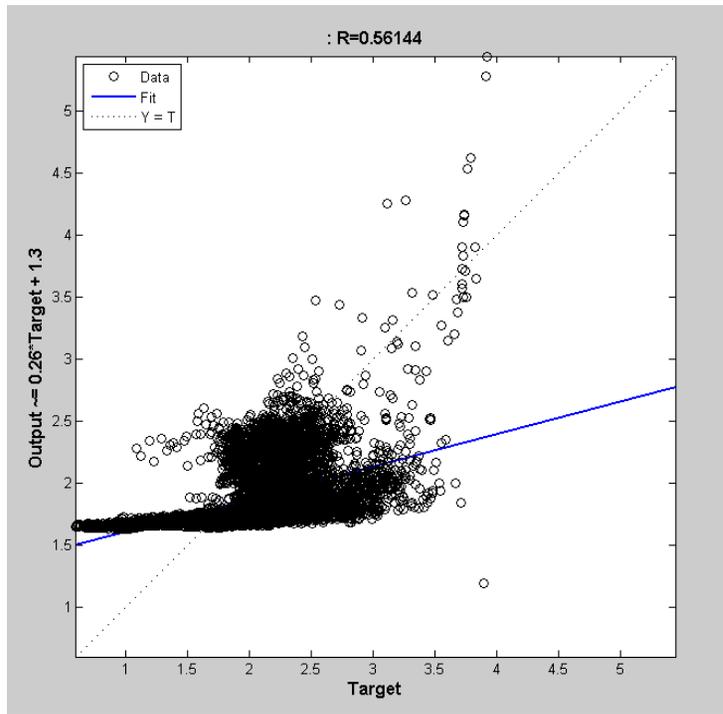

Figure 4.5: Regression plot for resistivity logs

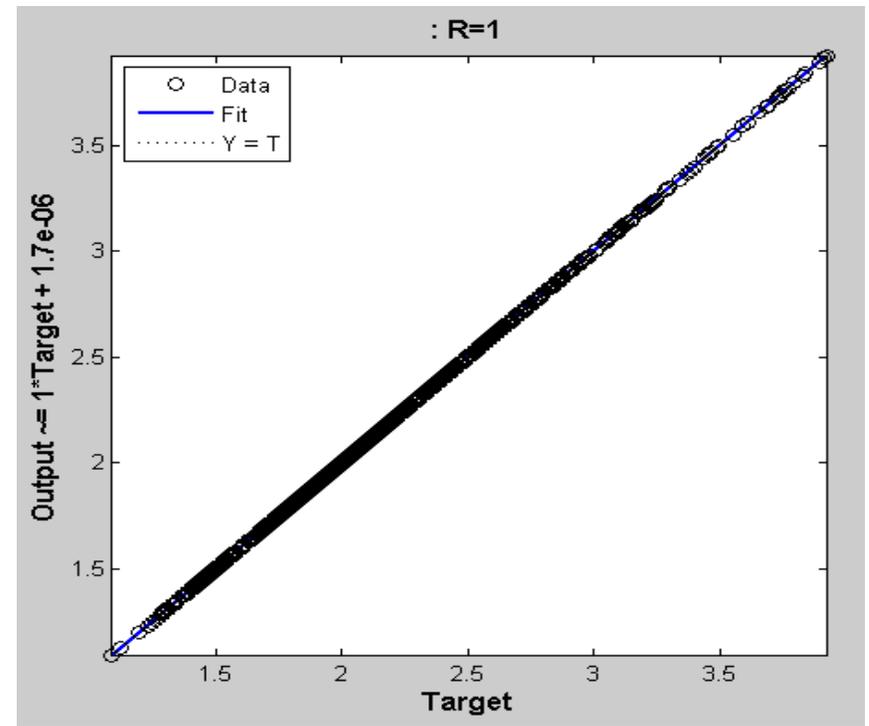

**Figure 4.6: The Sonic Log Effect**

In this Study, 8 well logs together with depth from two different wells were used. Shear wave velocity data were generated from the sonic logs using the Castagna Equation. Firstly, well logs which have the highest correlation with shear wave velocity were determined and utilized in regression and Artificial neural network. The network was a back propagation neural network with 1 or 3 neurons in the hidden layer, with tangent sigmoid activation function in the hidden layer and purelin activation function in the output layer.

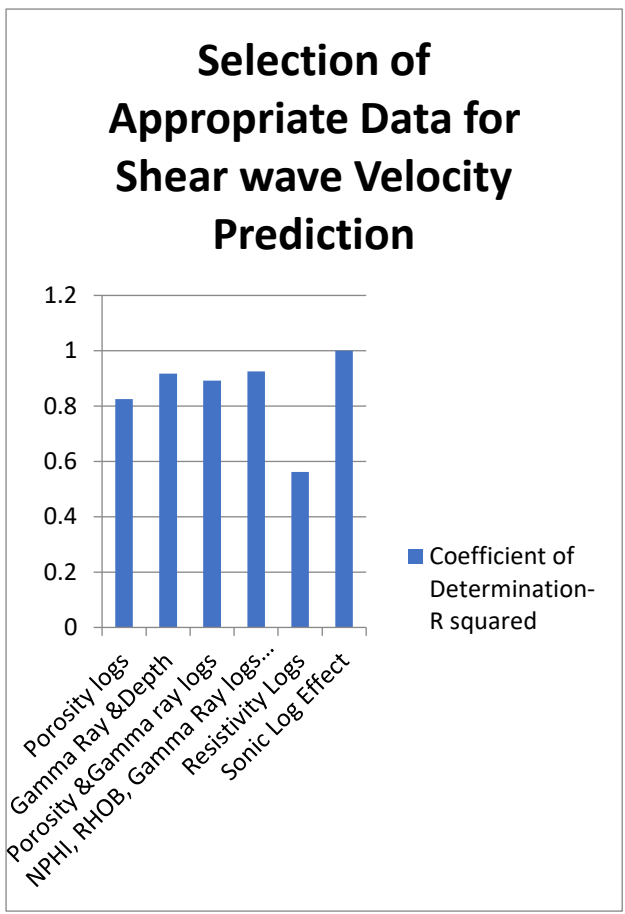

**Figure 4.7: Graph of R squared value of various combinations of well logs in predicting shear wave velocity.**

The figure above shows a comparison of the co-efficient of determination ($R^2$) for a combination of the logs considered for the determination of shear wave velocity. It is observed, from the graph, that resistivity logs have the lowest $R^2$ value among the logs. This implies that resistivity has little correlation with shear wave velocity, and hence, resistivity logs cannot be used in the estimation of shear wave velocity.

Porosity logs are also seen to have a relatively high co-efficient of determination ($R^2$). This implies that porosity logs are a key parameter in the determination of shear wave velocity. Depth and gamma ray logs show a quite high $R^2$ value. This indicates the significant influence of depth on the porosity of the formation and the dependence of shear wave velocity on the lithology of the formation (gamma ray log)

A combination of porosity, gamma ray and depth logs show the highest co-efficient of determination ($R^2$). This implies that the best way to estimate shear wave velocity is by using not just one parameter as in the case of porosity logs, but rather a combination of porosity, gamma ray logs for depth matching and depth logs.

## 4.2. COMPARISON OF METHODS

## 4.2 .1. Known Interval

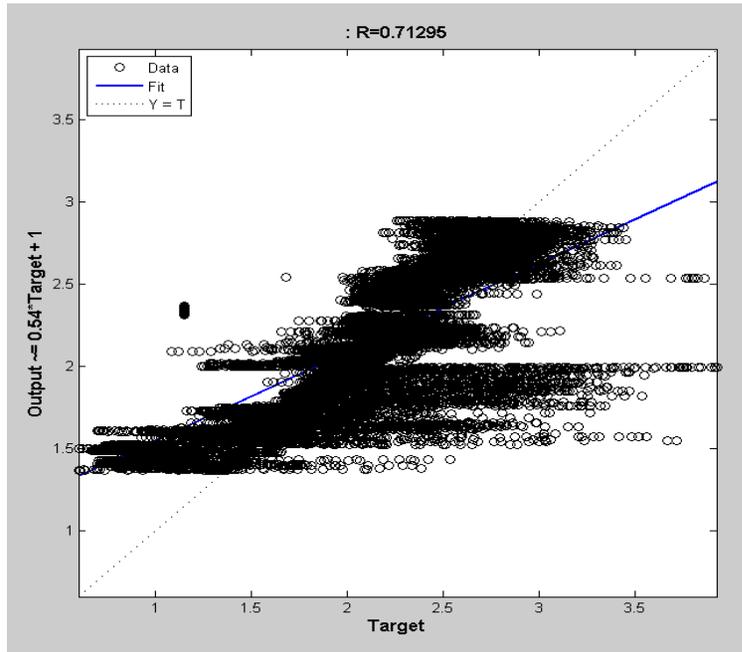

Figure 4.8: Regression graph using Single Variable Linear Regression

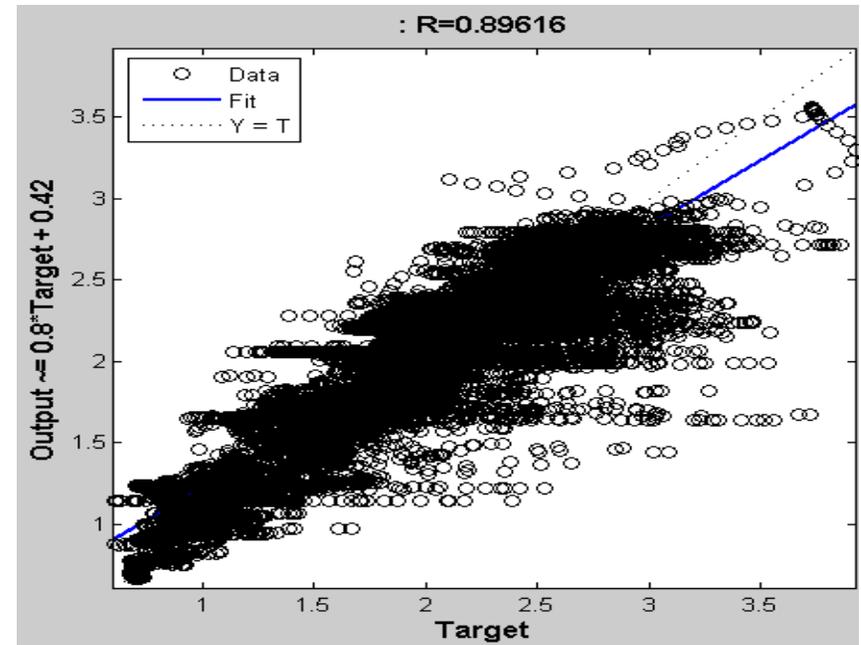

**Figure 4.9: Regression graph using ANN-Single Variable**

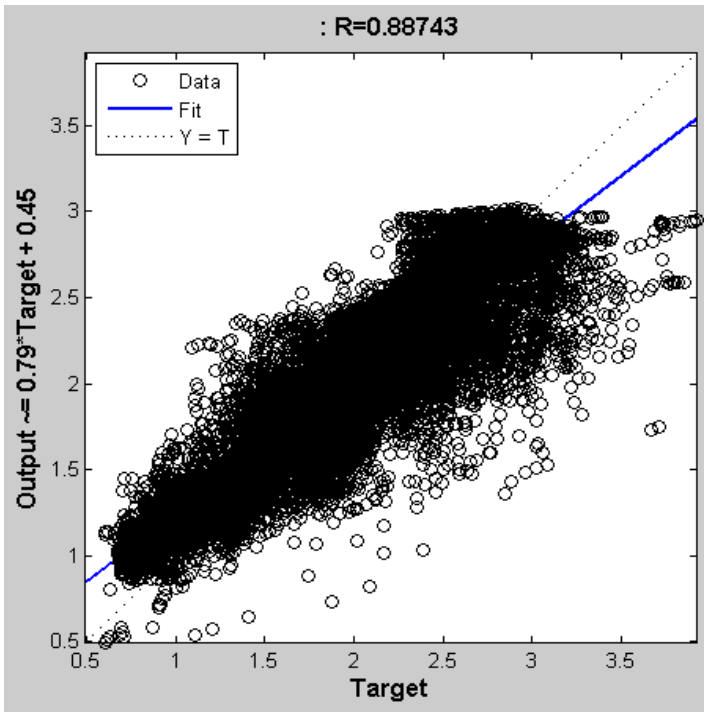

Figure 4.10: Regression graph using Multiple Linear Regression

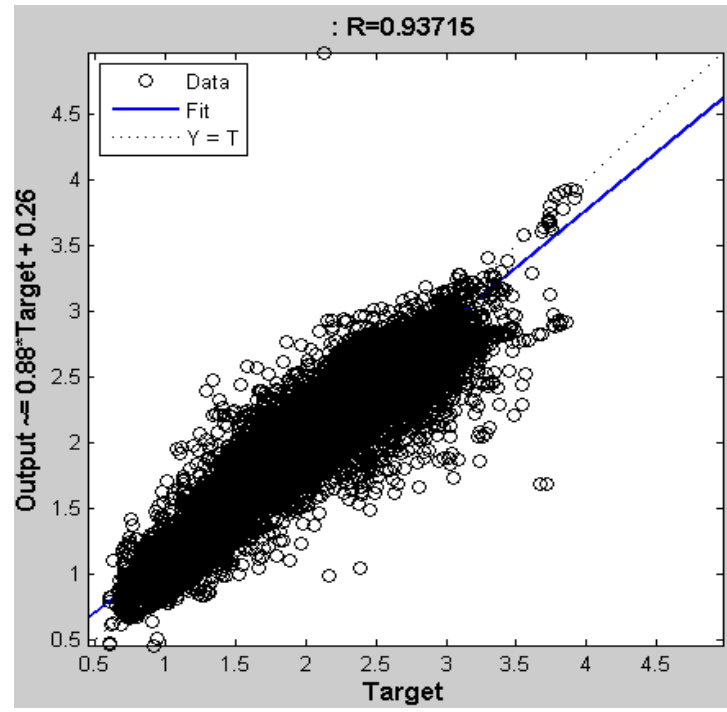

Figure 4.11: Regression graph using ANN- Multiple Variable

34

## 4.2.2. Unknown Interval in Same Well

Figure 4.12: Regression graph using Single Variable Linear Regression

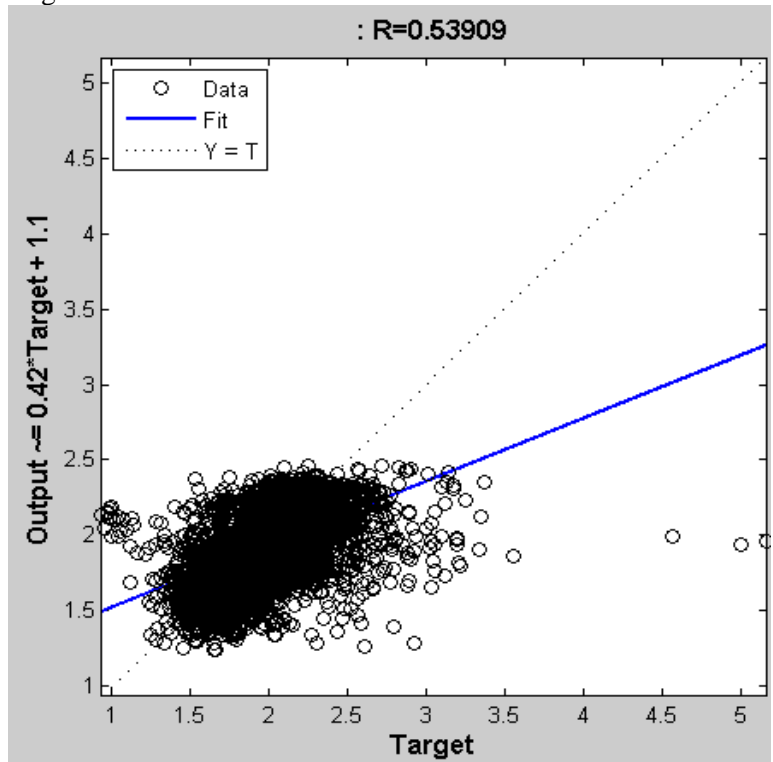

**Figure 4.12: Regression graph using Single Variable Linear Regression**

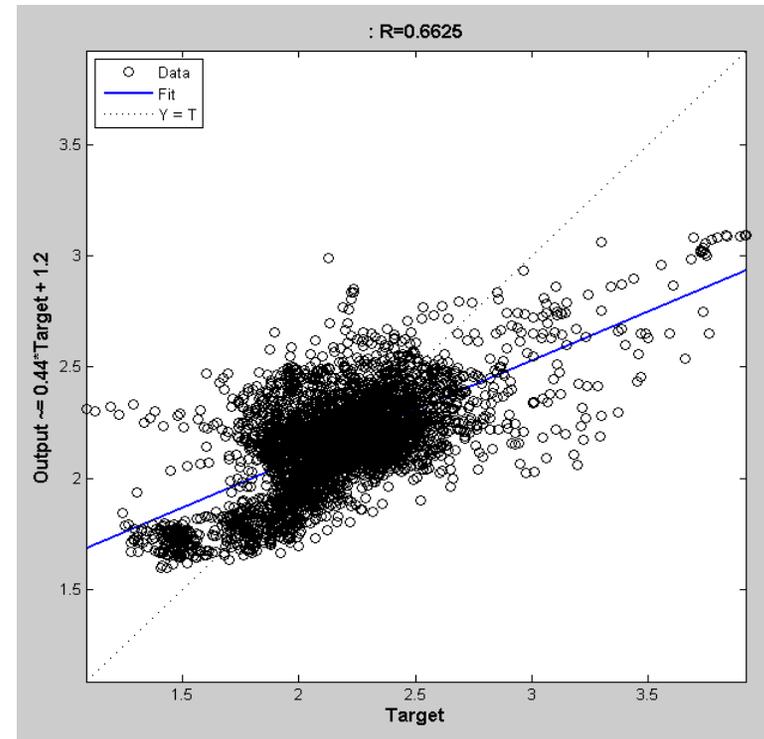

**Figure 4.13: Regression graph using ANN- Single Variable**

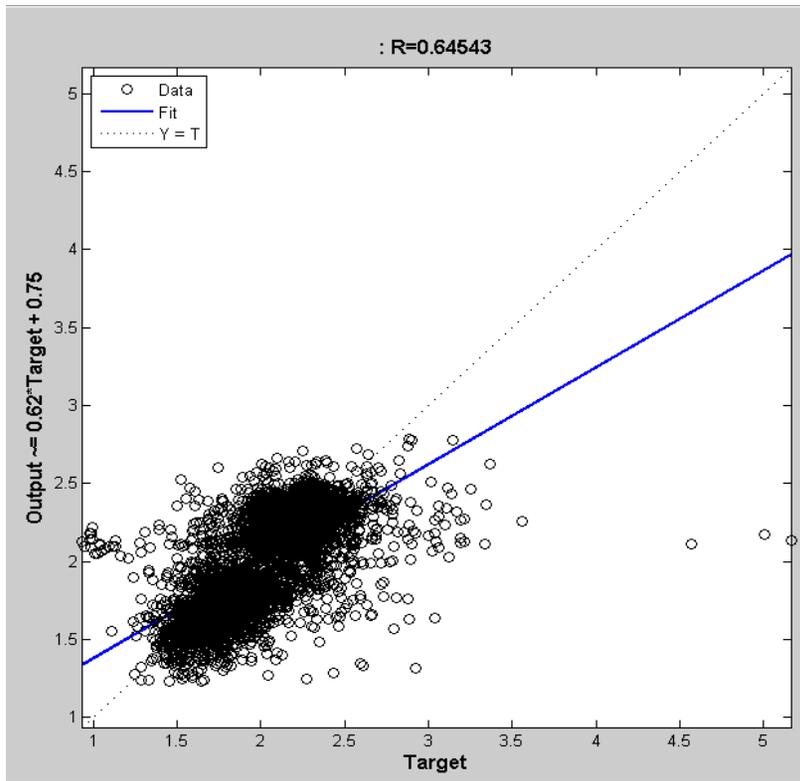

**Figure 4.14:Regression graph using Multiple Linear Regression**

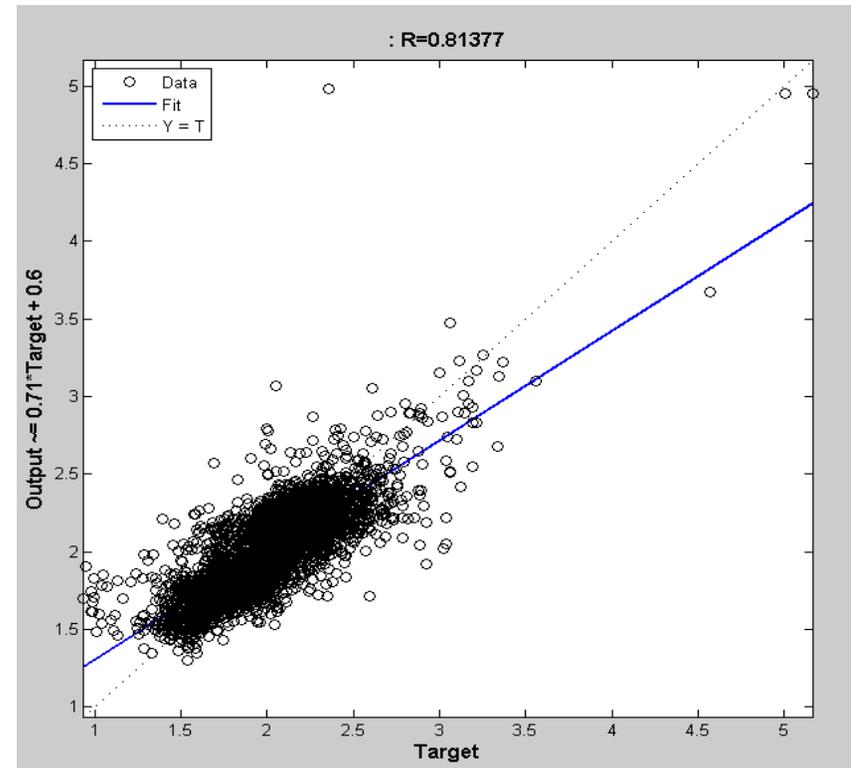

**Figure 4.15: Regression graph using ANN-Multiple Variable**

## 4.2.3. Unknown Interval in Different Well in Same field

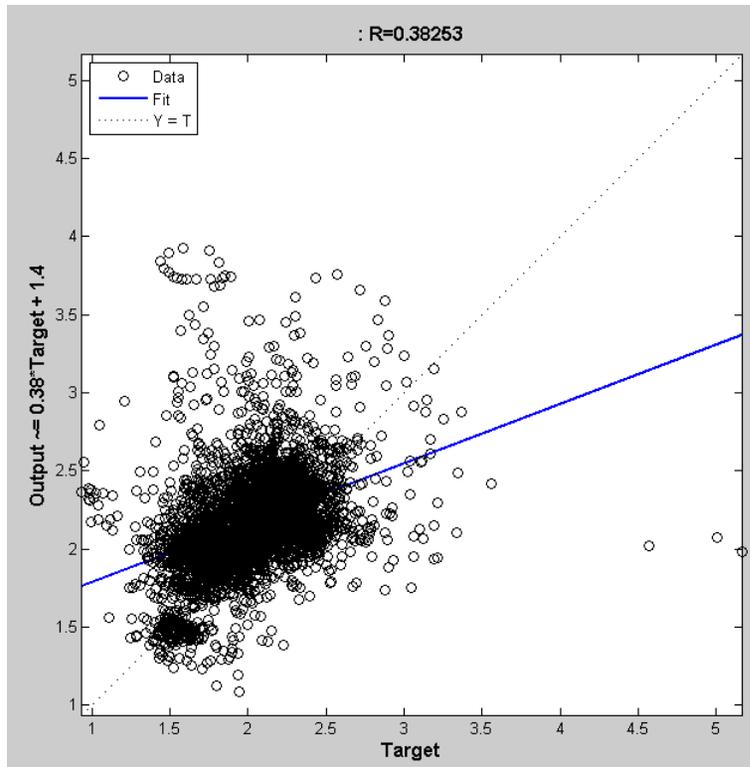

Figure 4.16: Regression graph using single variable Linear Regression

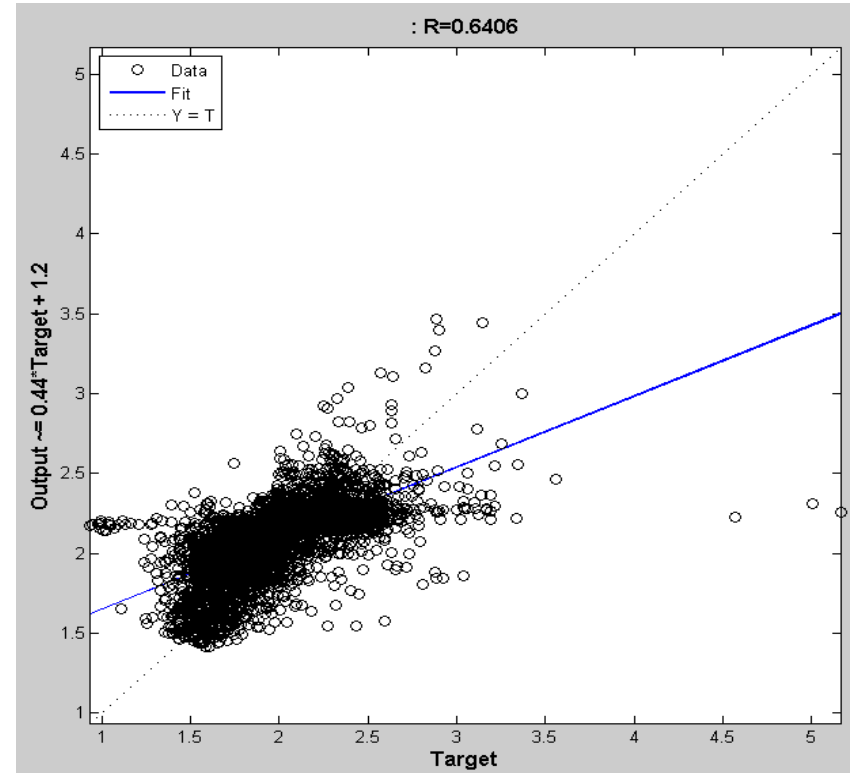

**Figure 4.17: Regression graph using ANN-single Variable**

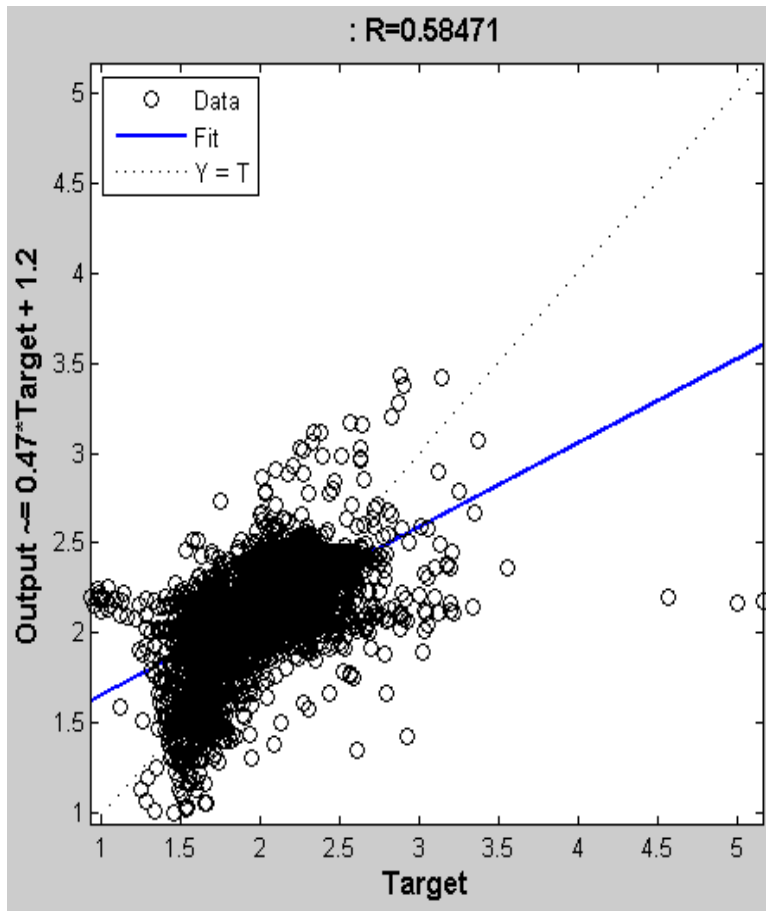

**Figure 4.18: Regression graph using Multiple Linear Regression**

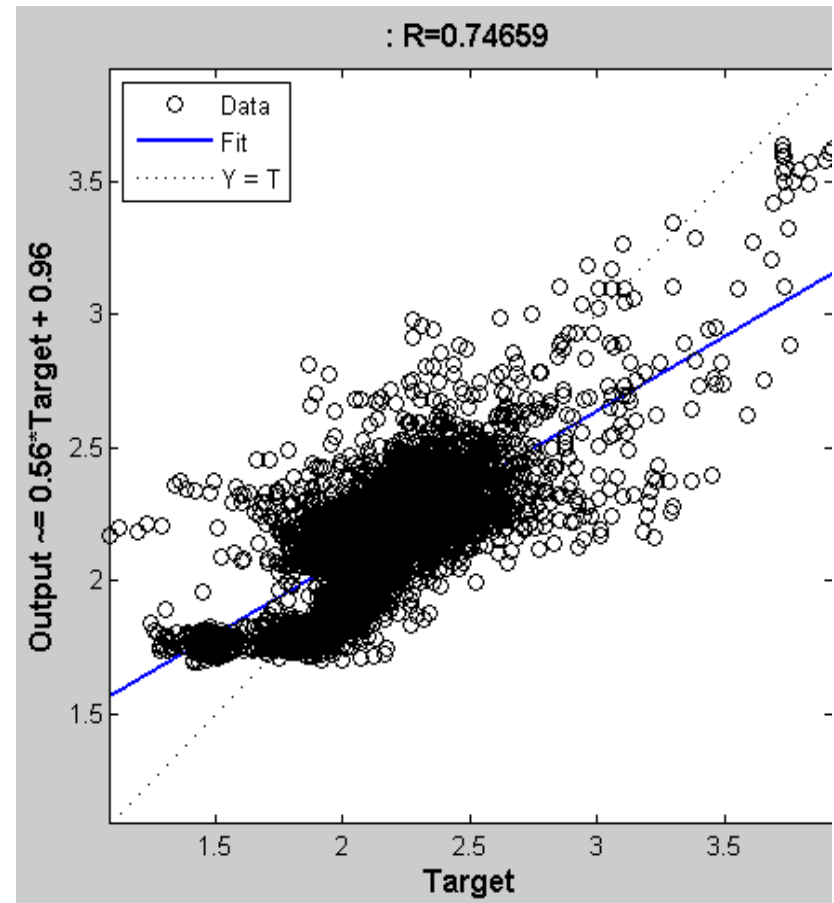

**Figure 4.19: Regression graph using ANN-Multiple Variable**

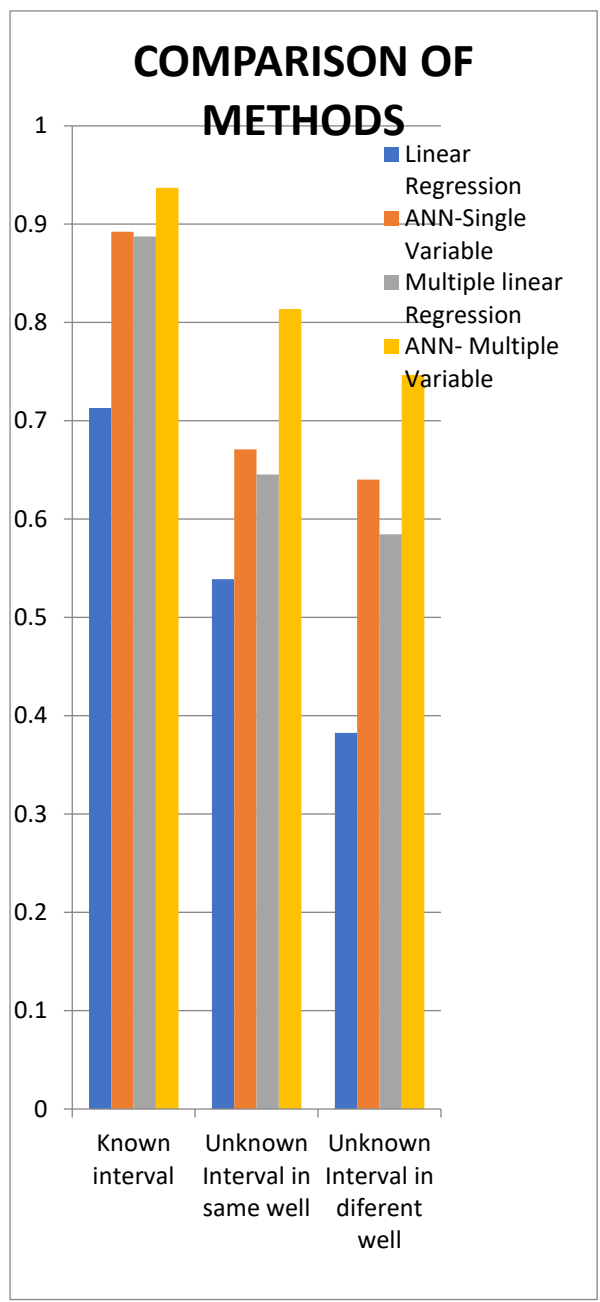

**Figure 4.20: Comparison of the R squared values of the various predictive methods under different conditions.**

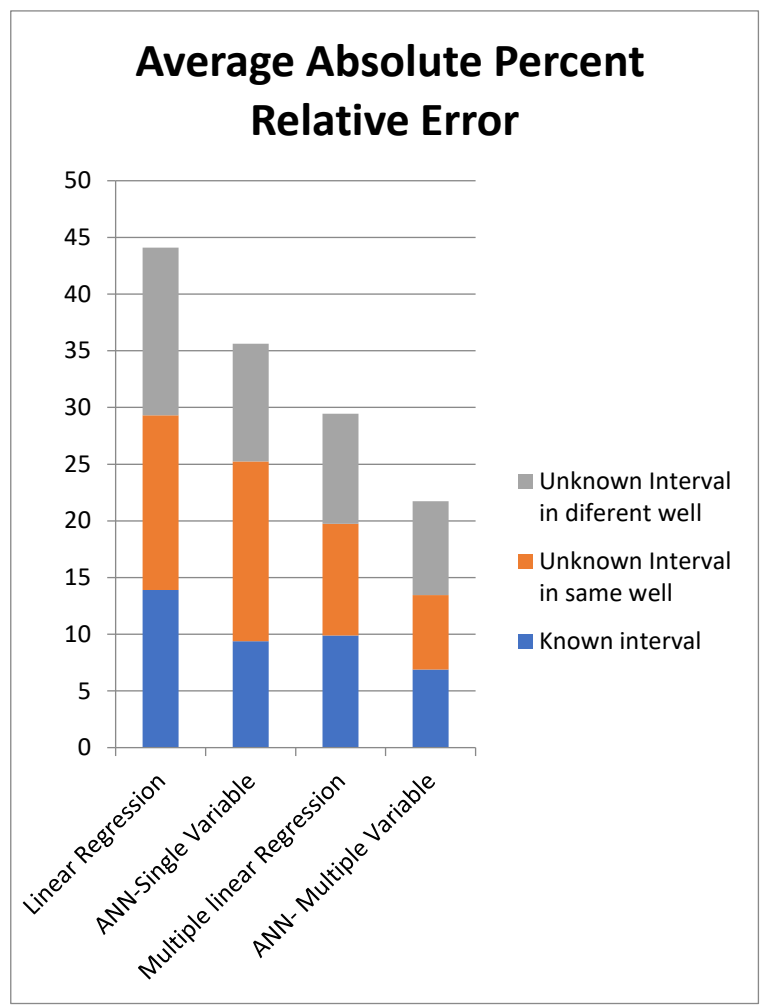

**Figure 4.21: Stacked column graph of the AAPRE associated with the various methods for each condition**

From the figure 4.13 above, the co-efficient of determination ($R^2$) for all the methods show quite high values, the single linear regression showing the least of the values. Multiple linear regression showed the next lowest, then ANN-single variable and the highest being ANN-multiple variable. This trend is seen for the unknown intervals in both wells under consideration for this case study.

There is a notable decrease in the $R^2$ for each of the methods used from the known interval to the unknown intervals, especially

39

for the ANN methods. This could be due to the fact that the networks were fed with the output for the known intervals, as opposed to the unknown intervals where the network had to predict the outputs. The regression methods gave good results in the known interval, but where it was applied to unknown intervals and new wells it usually faced problems. Such problems are avoided with the use of ANNs. ANNs have the ability to adapt data in the form of input-output patterns (dynamic regression) unlike the rigid regression methods.

Practically, the difference in the $R^2$ for each case could also be due to the fact that the two (training and testing) intervals have different properties. For an unknown interval in same well, the value of $R^2$ was closer to the $R^2$ of the known interval than that of the unknown interval in a different well. this shows that the properties of the two wells are much different and hence one well can hardly predict another.

Also, for all the intervals under consideration, the ANN- multiple variable showed the highest $R^2$ value among all the other methods. It could be inferred from the graph above that the ANN-multiple variable is the best method for the determination of shear wave velocity for unknown intervals in a well.

This inference is backed by the AAPRE of the various methods. Figure 4.14 shows the amount of deviation of the output of each method from the target. The method with least error is the ANN-Multiple variable and linear regression has the highest error.

## CONCLUSIONS AND RECCOMMENDATIONS

### 5.1 CONCLUSIONS

Multiple variable ANNs should be the most preferred method of estimation of shear wave velocity, especially at intervals in the well where the data is unavailable. This is because this method provides the least margin of error as compared to the other methods considered. Also, the accuracy of this method can be improved with repeated simulations and changes in the network, that is, it is more flexible and subject to change than the other methods. This method can also be used to estimate shear wave velocities in similar wells in the same formation.

In the absence of compressional wave data (sonic tool readings), using ANNs, with porosity, depth and gamma ray logs as inputs would be the most effective for the prediction of shear wave velocity.

### 5.2 RECCOMMENDATION

The unavailability of real field data on shear wave velocity made it difficult to determine shear wave velocity using its main dependent factor (sonic logs). In the future, shear wave velocity data should be made available to make the model much more credible and workable.